\documentclass{aa}
\usepackage{graphicx}
\usepackage{afterpage}
\usepackage{aalongtable}
\usepackage{epsfig}

\begin{document}
\newcommand{\vsini}{$v$sin $i$~}
\newcommand{\cmr}{~cm$^{-1}$~}
\newcommand{\Teff}{T$_{\rm eff}$~}
\newcommand{\mkm}{$\mu$m~}

\title{Spectral analysis of high resolution near-infrared spectra of ultra cool dwarfs}

\author{Yuri Lyubchik $^1$, Hugh R.A. Jones$^2$, Yakiv V. 
  Pavlenko$^{1,2}$, Eduardo Martin$^{3,4}$, Ian S. McLean$^5$, Lisa Prato$^6$, 
  Robert J. Barber$^7$, Jonathan Tennyson$^7$}

\offprints{Yuri Lyubchik}
\mail{lyu@mao.kiev.ua}

\institute{Main Astronomical Observatory of Academy of Sciences of
Ukraine, Zabolotnoho, 27, Kyiv, 03680 Ukraine
\and Center for Astrophysics Research, University of Hertfordshire,
College Lane, Hatfield, Hertfordshire AL10 9AB, UK
\and Instituto de Astrof\'\i sica de Canarias, La Laguna,
Tenerife 38200, Spain
\and University of Central Florida, Department of Physics, PO
Box 162385, Orlando, FL 32816, USA
\and Department of Physics and Astronomy, UCLA, Los Angeles, CA 90095
\and  Lowell Observatory, 1400 West Mars Hill Road, Flagstaff, AZ 86001
\and Department of Physics and Astronomy, University College London,
      Gower Street, London WC1E 6BT UK}

\date{Received ; accepted }

\authorrunning{Lyubchik Yu. et al.}
\titlerunning{Spectral analysis of high resolution near-IR spectra ...}

\abstract
{}
{We present an analysis of high resolution
spectra in the J band of five ultra cool dwarfs from M6 to L0.} 
{We make use of 
a new $ab$ $initio$ water vapour line list and existing
line lists of FeH and CrH for modelling the observations.}
{We find a good fit
for the Mn~I 12899.76\AA~ line. This feature is one of the few for which
we have a reliable oscillator strength.  Other atomic features are present but most of the observed features are
FeH and H$_2$O lines. While we 
are uncertain about the quality of many of the atomic line parameters,
the FeH and CrH line lists predict a number of features which are not
apparent in our observed spectra. We infer that the main limiting factor 
in our spectral analysis is the FeH and CrH molecular spectra.}
{}


\maketitle

\section{Introduction}

Observations show that M dwarfs are the most numerous population
in the solar neighbourhood. The total number of very low mass stars
and cool objects, which include L and T dwarfs, could be much
higher, e.g., from simulations Allen et al., \cite{allen} or from
inferences from nearby star counts Henry et al., \cite{henry}.

Molecular absorption  plays the definitive role in the appearance
of ultra cool dwarf spectra. Molecular bands 
govern the overall shape of the infrared observed spectra of cool
dwarfs. In spite of a lot of effort in measuring and computing
more accurate and full molecular line lists the situation is far from ideal.
This is caused by a lack of accurate oscillator strengths and
quantum-mechanical data for molecules
in the infrared region.

The situation is not ideal even for atomic lines which in comparison to molecular lines are 
relatively straightforward to obtain. Most of the
atomic lines in the infrared are formed by transitions between
upper levels of atoms which are difficult to compute with high
accuracy. As shown in Lyubchik et al., \cite{lyubchik}, strong atomic lines in the spectra 
of ultra cool dwarfs can be
identified even against a background of molecular absorption.
Comparison of these atomic lines with computed profiles can be
used as key diagnostics of ultra cool dwarf atmospheres. 

The peak of the energy
distribution for most cool and ultra cool dwarfs is located in or 
close to the J band.
Many authors have studied the J-band region of ultra cool dwarfs.
Kirkpatrick et al., \cite{kirkpatrick} and Jones et al., \cite{jones94},
\cite{jones96}  
provided identifications of spectral features and indicated it as a
productive spectral region to study.  
Molecular absorption in the J band is comparatively weak, 
so atomic lines of even intermediate strength are easily identified. 
The advent of the newest infrared spectrographs mean
that good quality high resolution spectra can be taken of brown dwarfs and
ultra cool dwarfs (e.g., see McLean et al., \cite{mclean07}, Zapatero Osorio et al.,
\cite{zapateroosorio} and spectra shown in our 
paper). However, despite these
advantages for spectral analysis the determination of the physical 
parameters has so far relied only on the strongest lines (e.g., 
see Leggett et al., \cite{leggett}).

\section{Observational data}

\begin{figure}[hbt]
\centerline{\includegraphics [width=65mm, height=85mm, angle=270]{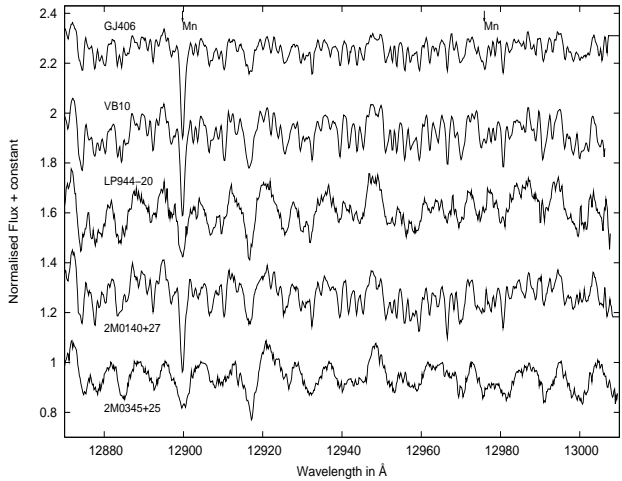}}
\caption{Spectral sequence of the targets in spectral type order. The positions
of 12899.76\AA~ and 12975.91\AA~ Mn~I lines are shown.}
\label{fig1}
\end{figure}

The observational data details and characteristics of the target stars 
are presented in Table~1. 
Hereafter we use the following designations: 2M0140+27 for 
2MASSW~J0140026+270150 and 2M0345+25 for 2MASP~J0345432+254023.
The spectra are shown in Fig. \ref{fig1}.

The spectra were obtained with NIRSPEC on Keck. The reduction of these
spectra was carried out according to standard procedures (McLean et al., 
\cite{mclean}), Prato et al., \cite{prato02a}, 
\cite{prato02b} and Zapatero Osorio et al., \cite{zapateroosorio}). 
Integration times ranged from
10 minutes to 1.5 hours depending on J-band magnitude.
For the faintest objects, the individual exposures were 600s,
for brighter objects - 300s, 120s, or 60s (see McLean et al., \cite{mclean07} and
Zapatero Osorio et al., \cite{zapateroosorio} for details). 

For all stars we use the 59 NIRSPEC echele order (12826--13015~\AA, see table 2 
in McLean et al., \cite{mclean07}). 
The resolving power of the spectra is R$\sim$20000 (it varies from 
R=17800 at 11480\AA~ to R=22700 at 13460\AA).
SNRs are in the range of $\sim$50 for 2M0345+25 to over 100 for GJ 406 and 
2M0140+27.

We note the presence of OH night sky emission lines though at
the relatively high-resolution (R$\sim$20000) used here 
they are not strong enough to saturate the detector.
If the OH line strengths have changed
in two subsequent exposures, the reduction software, REDSPEC,
removes any residual OH night sky lines (see McLean et al., \cite{mclean07} for
details).

Our targets are stars with non-zero radial velocities (e.g., see Martin et al.,
\cite{martin06}). 
We shifted the spectra to zero velocity using the Mn~I 12899\AA~ line which is a 
prominent feature in all spectra.

\section{Procedure}

\subsection{Synthetic spectra modelling}

Synthetic spectra are computed using the WITA6 programme
(Pavlenko, \cite{pavlenko00a}) for
NextGen model structures (Hauschildt et al., \cite{hauschildt}).
Calculations are carried out under the assumption of local
thermodynamic equilibrium, hydrostatic equilibrium, in the absence
of sources and sinks of energy. 
The atomic line list used for our spectral modeling
and line identification is taken from the VALD database (Kupka et
al., \cite{kupka}). The solar abundances reported by Anders \&
Grevesse, \cite{anders} are used in calculations. 
All details of other input parameters are described by Pavlenko
et al., \cite{pavlenko00b}. 

To compute the synthetic spectra we used molecular line lists from
different sources discussed below.
The relative contributions of these molecules to the formation of
the spectra are shown in Fig. \ref{fig2}.
Theoretical spectra are
computed with a wavelength step of 0.01\AA~ and convolved
with Gaussians to match the instrumental broadening. 
We adopt a FWHM of 1.2\AA. 
The rotational broadening of spectral lines is implemented following
Gray, \cite{gray}.

\subsubsection{Water vapour line lists}

Water vapour is the main contributor to the opacity across most of the
infrared for late type M dwarfs and brown dwarfs.  
There are several
H$_2$O line lists which are used in computations of
synthetic spectra of dwarfs: HITRAN (Rothman et al., \cite{rothman}),
MT (Miller et al., \cite{miller}), AMES (Partridge \&
Schwenke, \cite{partridge}), SCAN (J$\o$rgensen et al., \cite{jorgensen}) and VT2 
(Viti et al., \cite{viti}). For details and comparisons of these line lists 
the reader is referred to Pavlenko, \cite{pavlenko02} and Jones et al., 
\cite{jones02}, \cite{jones03}, \cite{jones05}. 
In our computations we used the AMES and BT2 line lists. 
We give a short description of each line list.

The AMES (Partridge \& Schwenke, \cite{partridge}) water vapour line list was computed
using a potential energy surface (Murrell et al., \cite{murrell})
which was fitted to line frequencies with J$<$5,
given in the HITRAN (Rothman et al., \cite{rothman}, http://www.hitran.com ) database.
The line intensities are
calculated using an {\it ab initio} dipole moment surface.
It has helped enable a wide range of photometric and spectroscopic results.

The BT2 synthetic line list for H$_2$O
(Barber et al., \cite{barber}) was computed using a discrete 
variable representation implemented within
the DVR3D suite of programs (Tennyson et al., \cite{tennyson}) to calculate the
rotation-vibration spectra of triatomic molecules. 
BT2 also used a potential energy surface obtained by fitting 
to laboratory spectra (Shirin et al., 2003) and an {\it ab initio} dipole 
surface (Schwenke \& Partridge, \cite{schwenke00}).
It is the most complete water line list, comprising over 500 million transitions (65\% more
than any other list) and is the most accurate one (over 90\% of all
known experimental energy levels are within 0.3\cmr of the BT2 values).
Tests by Barber et al., \cite{barber} show
that at 3000~K, BT2 includes 99.2\% of the water opacity in the J band
region, which represents a halving of the (1.6\%) opacity missing
in the previously best-available list, AMES (Partridge \& Schwenke, \cite{partridge}). 
Nonetheless our analysis is concerned 
with objects where the temperatures of the line forming regions are less than 3000~K and so 
it is perhaps unsurprising that our analysis is insensitive to the differences in synthetic 
spectra computed using AMES and BT2 H$_2$O line lists. Here we  show the
spectra computed using the BT2 line list. 

\subsubsection{Other molecules}

Molecules of FeH and CrH are also important opacity sources 
in the spectra of ultra cool dwarfs. In our calculations 
we used line lists of FeH (Dulick et al., \cite{dulick}) and CrH (Burrows 
et al., \cite{burrows}).

The line list of FeH (Dulick et al., \cite{dulick}) pertains to the
$F^4\Delta_i - X^4\Delta_i$ transitions using
the spectroscopic constants of rotational levels
(v = 0, 1, 2) of the FeH X and F states. Dulick et al. extrapolate for values 
v = 3, 4 and J-values up to 50.5. The line
list of FeH we use then consists of experimental and extrapolated term values
for the 25 vibrational bands with v$\le$4.
In this line list $^{54}$Fe, $^{57}$Fe and $^{58}$Fe along with the 
most abundant isotope $^{56}$Fe (91.7\% of iron abundance) are 
accounted for.

The CrH line list (Burrows et al., \cite{burrows}) contains information on
lines of 12 bands of the $A^6\Sigma^+ - X^6\Sigma^+$ electronic
system and includes ground state term values
for vibrational levels v = 0, 1, 2 up to J = 39.5. For terms with v=3 and high J, 
Burrows et al. extrapolate from Bauschlicher et al., \cite{bauschlicher}.
The most abundant $^{52}$Cr isotope (83.3\% of chromium abundance),
$^{50}$Cr, $^{53}$Cr and $^{54}$Cr are also included in CrH line list.

We include TiO absorption (Plez, \cite{plez}) in our spectral analysis though TiO 
is not a significant contributor 
at these wavelengths (see Fig. \ref{fig2}). We note, that another TiO line list
by Schwenke, \cite{schwenke98} may be used to account for TiO absorption. 
However, Pavlenko et al., \cite{pavlenko06} indicates that both sources 
will provide similar opacity over this spectral region .

\section{Results}

\begin{figure}[hbt]
\centerline{\includegraphics [width=65mm, height=85mm, angle=270]{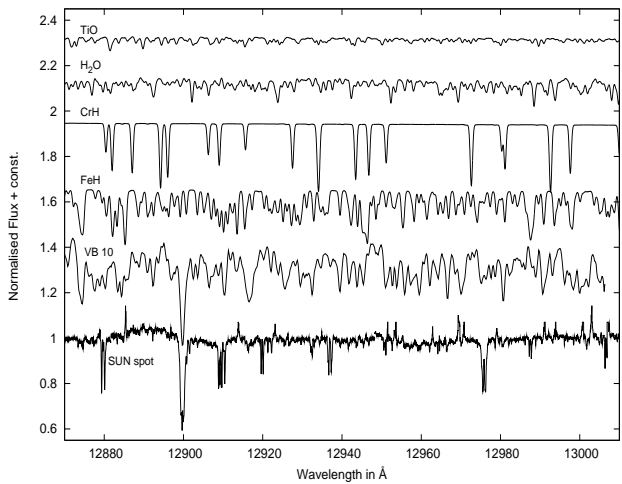}}
\caption{Spectral sequence of observed spectra of Solar umbra, VB10 and
synthetic calculations for FeH, CrH, H$_2$O (BT2 line list) and TiO
molecules across the region of interest.}
\label{fig2}
\end{figure}

We compute the individual contributions of the atomic and molecular
species for the wavelength region 12800-13010\AA~ (Fig. \ref{fig2}). While TiO 
and H$_2$O do not show strong lines in the  
region, their numerous absorption lines (especially H$_2$O) form the pseudo-continuum 
above 2500K. As the effective temperature drops below 2500K the contribution to the total 
opacity by FeH and CrH becomes dominant. 

\begin{figure}[hbt]
\centerline{\includegraphics [width=85mm, height=110mm, angle=0]{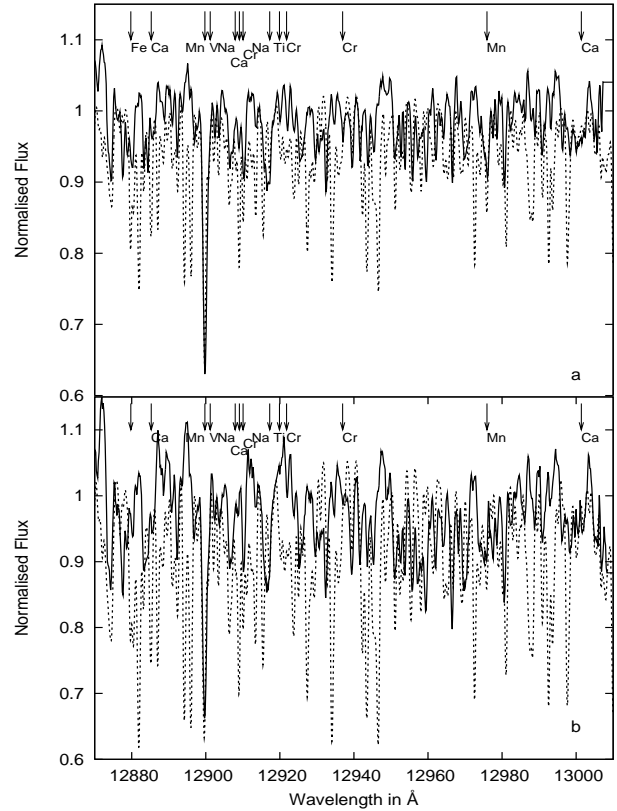}}
\caption{(a) The observed GJ406(M6~V) and (b) 2M0140+27(M8.5~V)
spectra (solid line) compared to the 2800/5.0/0.0 and 2500/5.0/0.0
synthetic spectra, computed using molecular line lists of 
H$_2$O(BT2), TiO, FeH, CrH (dotted line).}
\label{fig3}
\end{figure}

In Fig. \ref{fig3} we show the comparison of synthetic spectra 
2800/5.0/0.0 (\Teff=2800~K,
log~g = 5.0 and [M/H] = 0.0) and 2500/5.0/0.0, computed using TiO,
CrH, FeH and H$_2$O (BT2) molecular line lists, with the observed GJ406 
(panel a) 
and 2M0140+27 (panel b) spectra.

CrH lines can be identified in the observed spectra (see Table 2).
However, analysis of the observed CrH spectrum is very problematic 
due to the high multiplicity of the upper electronic state of the A-X 
transition of CrH 
(Burrows et al., \cite{burrows}). From comparison of Figs. \ref{fig3} 
(a), (b), \ref{fig4} and \ref{fig6} we find the use of the CrH opacity 
causes a poorer fit both for GJ406 and 2M0140+27. 
From Fig. \ref{fig3} onward we show the synthetic spectra computed without
the CrH line list.

We compute a set of synthetic spectra for model atmospheres of
different effective
temperatures appropriate to our cool dwarfs. Assuming our
observed stars are dwarfs, we adopt log~g=5.0 for all
computations except for LP944-20 where we used a value of
log~g=4.5 (Pavlenko et al., \cite{pavlenko06b} ). We find the differences between 
synthetic spectra computed for models of $\Delta$log~g=$\pm$0.25 and 
[M/H] = $\pm$ 0.2 are rather marginal. Given the lack of evidence for non-Solar 
metallicities in the literature for our target stars, we assume Solar metallicities.

Our choices of effective temperatures are based on the spectral type 
estimations by different authors (see Table 1).
Golimowski et al., \cite{golimowski} provides effective temperature ranges
depending on the age of stars. For the hottest star from our
program list GJ~406 and the coolest star 2M0345+25 Golimowski et al. give
the ranges of \Teff=2650 -- 2900~K and \Teff=2000 -- 2350~K, for assumed ages of 
0.1 -- 10 Gyr and known parallax uncertainties. 
The fit of the strong Mn features is consistent with the mid-point of these effective 
temperatures to within 100K.
 
\subsection{GJ 406 \& VB10}

\begin{figure}[hbt]
\centerline{\includegraphics [width=65mm, height=85mm,angle=270]{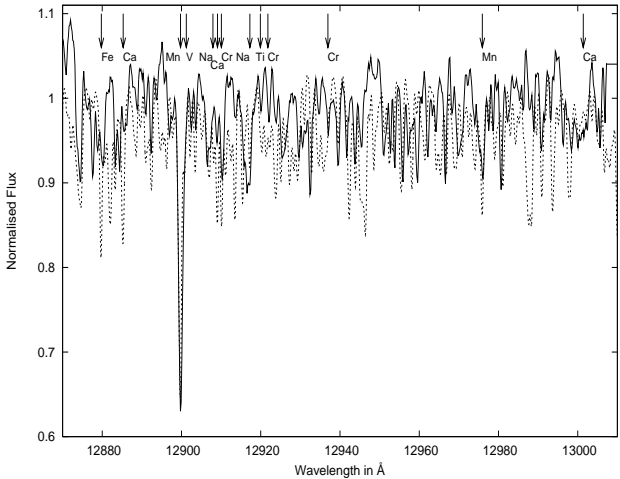}} 
\caption{The observed spectrum of GJ406~(M6~V) (solid line) compared to
a 2800/5.0/0.0, \vsini = 3km/s synthetic spectrum using molecular line lists of
H$_2$O(BT2), TiO and FeH (dotted line). } 
\label{fig4} 
\end{figure}

\begin{figure}[hbt]
\centerline{\includegraphics [width=65mm, height=85mm,angle=270]{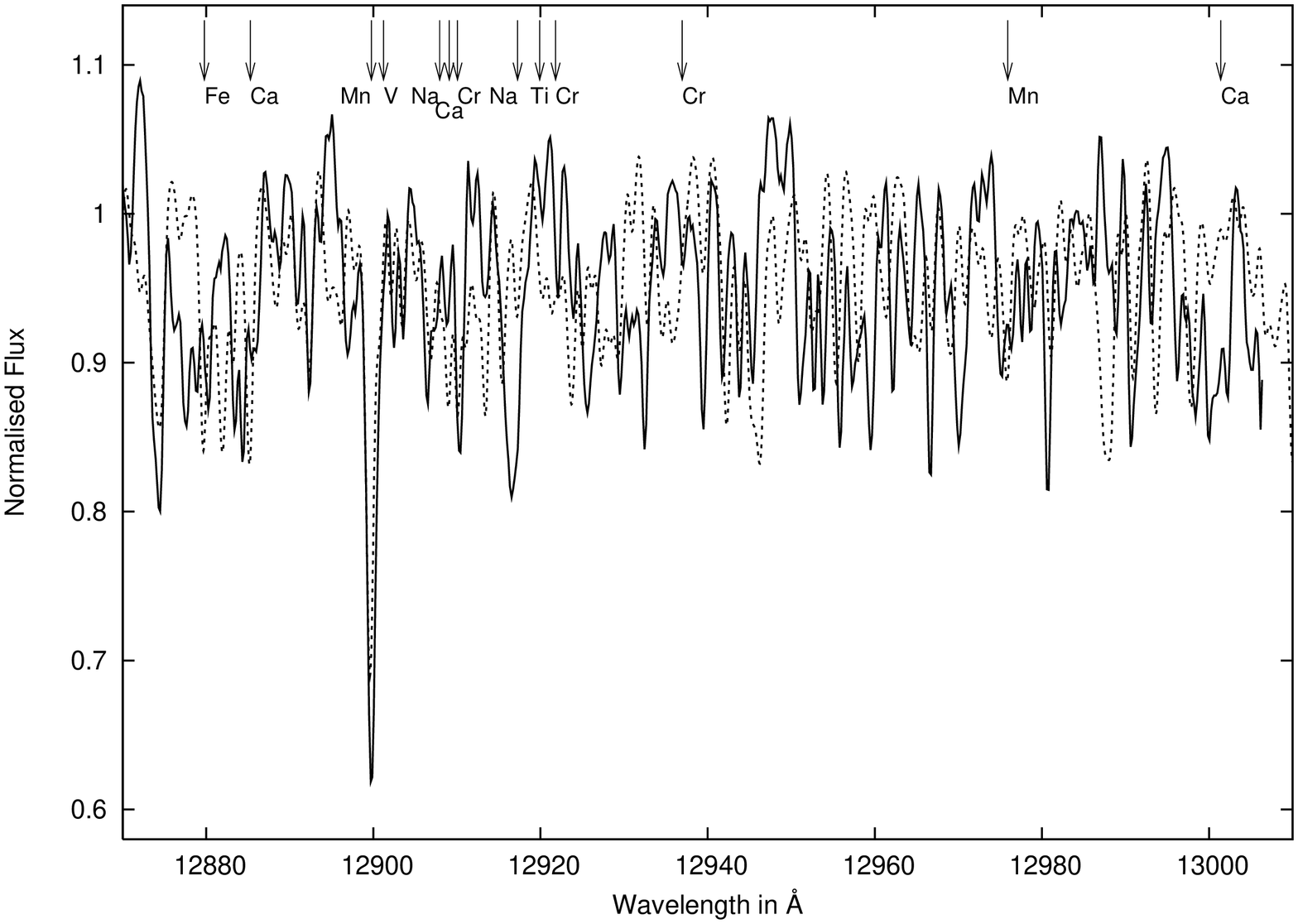}} 
\caption{The observed VB~10~(M8~V) spectrum (solid line) compared to a 
2700/5.0/0.0, \vsini = 10km/s synthetic spectrum with molecular line lists of
H$_2$O (BT2), TiO and FeH (dotted line). } 
\label{fig5}
\end{figure}

A comparison of the observed 
spectrum of GJ~406 (M6~V) and a synthetic spectrum 2800/5.0/0.0 with \vsini $\sim$ 3 km/s 
(e.g., Mohanty \& Basri, \cite{mohanty}) is shown in Fig.\ref{fig4}.  Figure \ref{fig5} 
shows the compares the observed spectrum of VB~10 (M8~V) to a 
2700/5.0/0.0 synthetic one with \vsini $\sim$ 10 km/s (e.g., Jones et al., \cite{jones05}). 
For both 
observed spectra the positions of atomic lines in the synthetic spectra correspond to lines 
in the observed spectrum though intensities of computed and observed lines differ 
significantly for some lines. Many lines are blended by 
molecular lines of approximately the same strength. 
For example, the Fe~I 12879.76\AA~ line profile is described 
well in the Arcturus 
spectrum (see Lyubchik et al., \cite{lyubchik}) but could not be 
fitted in our spectra because of severe blending by molecular lines.

\subsection{2M0140+27}

\begin{figure}[hbt]
\centerline{\includegraphics [width=65mm, height=85mm,angle=270]{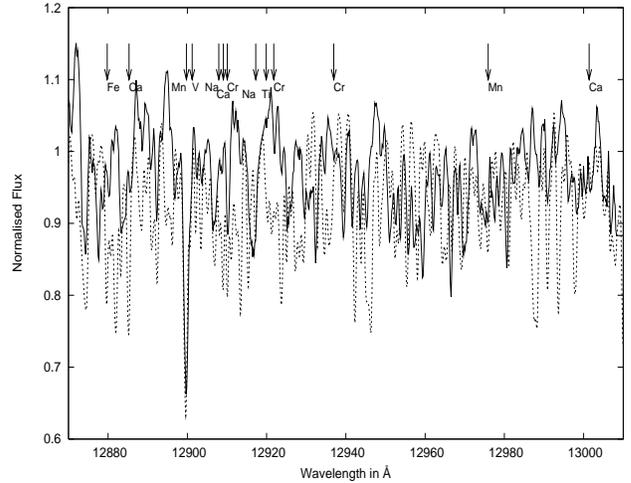}} 
\caption{The observed 2M0140+27~(M8.5~V)
spectrum is compared to the 2500/5.0/0/0, \vsini = 6.5 km/s synthetic spectrum with  
molecular line lists of H$_2$O(BT2), TiO, FeH (dotted line). } 
\label{fig6}
\end{figure}

In Fig.\ref{fig6} the observed spectrum of 2M0140+27 (M8.5~V) is compared 
to a synthetic spectrum 2500/5.0/0.0 with \vsini = 6.5 km/s (based on
Reid et al., \cite{reid}). The Mn~I 
12899.76\AA~ line is well fit, though the fit to Fe and other
weak atomic lines is not good at all. The rotational 
velocity is similar to the observational resolution and thus does not significantly 
change the synthetic line profiles. Thus 2M0140+27 also provides a good template for the 
identification of lines and their analysis.

\subsection{LP944-20}

\begin{figure}[hbt]
\centerline{\includegraphics [width=65mm, height=85mm,angle=270]{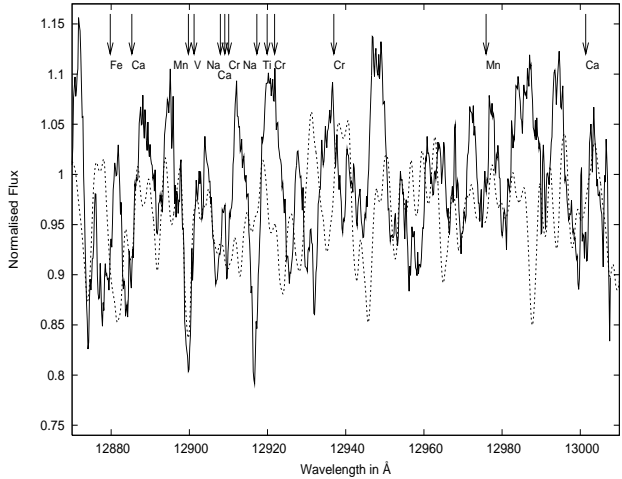}} 
\caption{The observed LP944-20~(M9~V)
spectrum (solid line) compared with the 2400/4.5/0.0, \vsini = 30 km/s synthetic 
spectrum with molecular line lists of H$_2$O(BT2), TiO, FeH (dotted line).} 
\label{fig7}
\end{figure}

LP944-20 is a fast rotating dwarf with \vsini $\sim$ 30 km/s (e.g., Tinney \& Reid, \cite{tinney}).
In Fig.\ref{fig7} we present observations of the brown dwarf
LP~944-20 (M9~V/M9.5~V) with a 2400/4.5/0.0 synthetic spectrum 
with \vsini = 30 km/s.
We find the large rotational broadening restricts the accuracy of our
analysis. Only stronger lines like Mn 12899.76\AA~ can be reliably 
identified in the observed spectrum. In general, the synthetic spectra
does not particularly resemble the observed one. This
may be partially explained by the effects of chromospheric-like structures,  vertical stratification in the outermost layers of LP944-20
(see Pavlenko et al., \cite{pavlenko06b}) or even abundance peculiarities.

The poor match of FeH absorption seems key.
In the synthetic spectrum we see some `features' created by the synthetic
FeH absorption at 12880, 12916, 12947,  12988 \AA, which are
absent in the observed spectrum. Features at these wavelengths can be 
recognized also in the VB10 spectrum. However, given that observed FeH
features should increase toward lower temperatures in late M-dwarfs
(Kirkpatrick et al., \cite{kirkpa}, Cushing et al., \cite{cushing}) we 
suspect these discrepancies are caused by FeH line list 
inaccuracies.

\subsection{2M0345+25}

\begin{figure}[hbt]
\centerline{\includegraphics [width=65mm, height=85mm,angle=270]{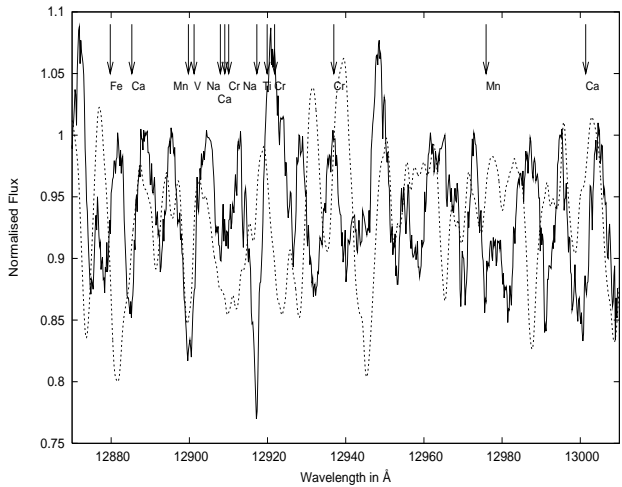}} 
\caption{The observed 2M0345+25~(L0) (solid line) spectrum is compared to the 
2200/5.0/0.0, \vsini = 45 km/s synthetic spectrum computed using molecular line lists 
of H$_2$O(BT2), TiO, FeH (dotted line).} 
\label{fig8}
\end{figure}

The lines in the spectrum of the L0 dwarf 2M0345+25 are very broad (see 
Fig.\ref{fig8}). We fit the observed 
spectrum of 2M0345+35 with a 2200/5.0/0.0 synthetic spectra (\vsini = 45 km/s). 
The \vsini value was determined by the synthetic spectra fit to the 
wings of Mn~I 12899.76\AA~ line in the observed spectrum.
The strong 
Mn~I 12899.76\AA~ line profile fits satisfactorily. Other atomic features, including Fe, 
seem to disappear under the strong pseudo-continuum formed by molecular lines. 
Again, as with LP944-20, the synthetic spectra including FeH predict a number of features in 
the synthetic spectrum which are not observed.

\subsection{Line identification}

Based on the observable spectra of three M-dwarfs which are less affected by rotation 
(GJ406, VB10 and 2M0140+27), we make 
identifications of absorption lines. These identifications are performed by comparison of 
each observed spectra with synthetic ones calculated for each molecular line list of the 
principal contributors to opacity in this region.

Results of these identifications are presented in
Table~2. We classify the absorption line-like features
in three groups based on their intensity: one asterisk (*) in columns 2, 3 and 4 of Table 2 corresponds to a 
mean "residual intensity" of the absorption line 
from 0.96 -- 0.99 (where a value of 1 corresponds to the
level of pseudo-continuum, not the real continuum level);
two asterisks (**) corresponds to a "residual intensity" between 0.91 -- 0.95; 
three asterisks (***) labels absorption lines deeper than 0.90. 
We also
check for these lines in the solar umbra spectrum (Wallace \& Livingston, \cite{wallace}, 
ftp://ftp.noao.edu/fts/spot1atl ) 
(column 5), asterisks mean the same as above.
Thus in column 6 of Table 2  we identify species, which form  
line-features in the observed spectrum, however, a number of synthetic lines 
are located at too-discrepant wavelengths or are too weak to make the observed spectral
features. In these cases we put a `?' to signify the uncertainty
in their identification.

As the next step in line identification we use lines, identified in
Table~2, with residual intensities larger than 0.9 (three asterisks in
columns 2,3 or 4), to find corresponding line-like features in the
spectra of the other two dwarfs: LP944-20 and 2M0345+25, with
highly broadened spectral features. These results are shown in Table~3.
Despite the large differences in intensity between observed and calculated lines for all 
features other than Mn some of the strong synthetic features are present in the observed 
data and can be identified with confidence. 

\subsection{Mn~I lines.}

Mn~I lines form clear features in the
spectra of ultra cool dwarfs in the J band, which is 
relatively free from distinctive molecular
absorption bands. In Lyubchik et al., \cite{lyubchik} we make detections 
using the VALD
line list (Kupka et al., \cite{kupka}; e.g. Kurucz's Mn~I data), 11 lines
of Mn~I lie in the region 12800 -- 14000\AA. 
Three of these lines are predicted to be rather strong in the
observed spectra of ultra cool dwarfs across the wavelength range of our observed spectra 
so we try to analyse these
lines, however, only the 12899.76\AA~ line is clearly detected in all
the observational spectra. The others are camouflaged and blended by
molecular lines. The quality of three spectra allow us to measure
the "pseudo equivalent width" (pEW) of the 12899.76\AA~ line. 
We note that these values are not the real equivalent width 
because of uncertain 
continuum level determination (Pavlenko, \cite{pavlenko1997}).
Using DECH20 package (Galazutdinov, \cite{galazutdinov}) we 
determine  
pEW(Mn~I)=501~m\AA, 500~m\AA~, and 502~m\AA~ for GJ406, VB10 and
2M0140+27, respectively. 
The accuracy of pseudo equivalent width determinations 
is $\sim$5\%.

Mel$\grave e$ndez, \cite{melendez} and Holt et al., \cite{holt}
indicate that the profiles of Mn lines and abundance determination 
depend on the hyperfine structure (HFS) and it is thus important to
consider HFS for accurate abundance analysis.
In Table~4 the available parameters of the Mn~I lines used in
computations are presented. Importantly an experimental log~gf
value of the 12899.76\AA~ Mn~I line obtained by Imperial
College group (UK) is close to the VALD value (Blackwell-Whitehead et al., 
private communication). 

\begin{figure}[h]
\centerline{\includegraphics [width=55mm, height=90mm,angle=270]{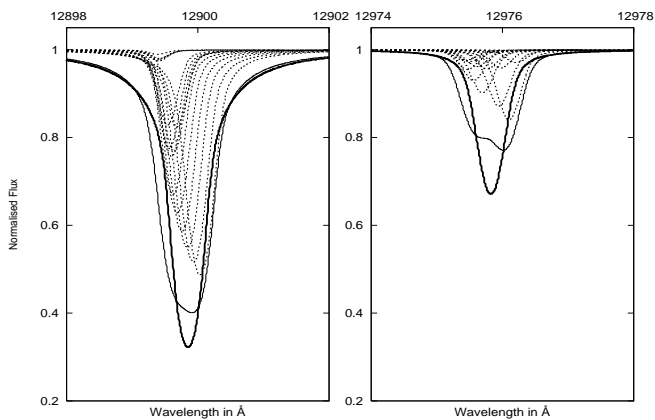}}
\caption{Computations of Mn~I line profiles with and without hyperfine
splitting:
the thick solid line - VALD log~gf for Mn~I, thin solid line - computations 
with all hyper-fine components (Mel$\grave e$ndez, \cite{melendez}),
dashed lines - computations for each hyper-fine component of Mn~I line.}
\label{fig9}
\end{figure}

The synthetic spectra to examine the HFS of Mn~I are computed using NextGen model 
atmospheres (Hauschildt et al., \cite{hauschildt}) with parameters 
\Teff=2800~K, log~g=5.0, [M/H]=0.0. The data on wavelengths and relative intensities of 
hyperfine
components for computations are taken from Mel$\grave e$ndez, \cite{melendez}. 
From Fig. \ref{fig9} 
it is seen that line profiles computed using the VALD data and the
average of HFS curve do not differ in a drastic way for Mn~I 12899.76\AA~ and
12975.91\AA~ lines.
However, our problems with making reliable line identifications mean
that before HFS effects are reliably diagnosed at these resolutions it 
will be necessary to get a more reliable fit to the bulk of weaker 
features which are providing the pseudo-continuum.

\section{Discussion}
In the case of the strong Mn lines, where we have reliable knowledge of the oscillator 
strengths we find satisfactory fits. This indicates that the overall level of opacity is 
reasonably well described.
However, apart from the Mn lines we have found a rather poor match between our observations 
and synthetic spectra across the 12810--13010\AA~ region.
We are not aware of high resolution spectral analysis using recent water line lists in this 
region, however, the previous good 
fits have been obtained with observations at longer wavelengths (e.g., Jones et al., 
\cite{jones05}) and our knowledge of the line list construction mean that we have confidence 
that water is reasonably well described over this region. Furthermore, 
in our calculations we used two different
line lists of H$_2$O: AMES (Partridge \& Schwenke, \cite{partridge}) and BT2 (Barber et al., 
\cite{barber}) and our inter-comparison of these line lists indicate that differences 
between them in the 12810--13010\AA~ are not significant for this analysis. According to our 
opacity calculations molecular opacities other than water, namely FeH and CrH, are important 
for an adequate modelling of the 12810--13010\AA~ region. 
However, the features predicted for these FeH and CrH line lists are not clearly evident. 
We thus suspect that the lack of overall match is due primarily
to problems with these line lists.

With the exception of the Mn line our fits between observed and 
synthetic spectra is poor. Nonetheless, the quality of other atomic opacities are difficult 
to judge since they are too weak to reliably distinguish from the pseudo-continuum. 
Thus the poor fits of the weak atomic lines cannot be considered as a 
product of poorly known atomic parameters. Furthermore, across all of the temperature range 
considered 
for our observed spectra we expect and observe blends formed by atomic and molecular lines. 
It thus seems that further computations and/or measurements of 
molecular line lists for FeH and CrH will be necessary in order
to carry out J band spectral synthesis of ultra cool dwarfs. 
Without such line lists spectral analysis of atomic features will be limited to a very small 
number of strong lines. 

In our analysis we did not use
dusty model atmospheres. Many authors have stressed that dust
is important for atmospheric temperatures 
$<$2700K (e.g., Jones \& Tsuji, \cite{jones97}). 
However, comparisons of synthetic spectra computations using
dusty models (Allard, \cite{allard} ) and non-dusty ones (Hauschildt et al., 
\cite{hauschildt}) show 
that dust effects are considerably lower than uncertainties in line lists of molecules.
Furthermore, dust opacity is modelled to produce a smooth opacity 
in a relatively high resolution study such as this. 
If dust is the dominant opacity 
across this region we would not anticipate to be able to fit the Mn line at 12899.76\AA.

\section{Acknowledgements}

We would like to thank the authors of the PHOENIX model atmospheres and the Sunspot 
umbra atlas for making their data available through their ftp sites.
The NSO/Kitt Peak FTS data used here were produced by NFS/NOAO.
The authors thank the Royal Society
and PPARC for support of travel and experimental work.

\newpage
\clearpage

\newpage
\begin{table}[ht]
\caption[]{List of programme stars including their spectral type 
(based on optical spectroscopy),
effective temperatures  and vsin$i$ (with references in brackets).
}
\begin{tabular}{cccc}
\hline\hline
Name                               & Spectral type           & \Teff     & vsin$i$, km/s \\
\hline
GJ 406 (LHS 36)                    & M6~V (K91)            & 2800K (B00) & 3 (MB03)\\
VB 10 (LHS 474)                    & M8~V (K91)            & 2700K (S96) & $\sim$10 (J05)\\
2MASSW J0140026+270150             & M8.5~V (Gi00)         & 2500K       & 6.5 (R02)\\
LP 944-20 (2MASSW J0339352-352544) & M9~V(K99)/M9.5~V(R02) & 2400K (B00) & $\sim$30 (TR98)\\
2MASP J0345432+254023              & L0 (K99)              & 2200K (B00) & 45 (*)\\
\hline
\multicolumn{4}{l}{B00 - Basri et al. 2000; Gi00 - Gizis et.al. 2000; J05 - Jones et al. 2005;}\\
\multicolumn{4}{l}{K91 - Kirkpatrick et al. 1991; K99 - Kirkpatrick et al. 1999; MB03 - Mohanty\&Basri 2003}\\
\multicolumn{4}{l}{R02 - Reid et al. 2002; S96 - Schweitzer et al. 1996; TR98 - Tinney\&Reid 1998}\\
\multicolumn{4}{l}{(*) determined using spectra fitting (see S.4.4)}\\
\end{tabular}
\end{table}

\newpage
{\clearpage
\begin{longtable}{cccccl}
\caption{Line identifications and strengths for the programme stars GJ406, VB10 and 2M0140+27 are tabulated.}\\
\hline\hline
WL$\pm$0.1\AA, air &  GJ406  & VB10 & 2M0140+27 & Sunspot & Notes \\
\hline
\endfirsthead
\caption{continued.}\\
\hline\hline
WL$\pm$0.1\AA, air &  GJ406  & VB10 & 2M0140+27 & Sunspot & Notes \\
\hline
\endhead
\hline
\endfoot
12874.4   & **  & *** & *** & **         & FeH, H$_2$O(?) \\
12877.7   & **  & *** & *** & * (wk)     & FeH(?), TiO(wk,?), H$_2$O(?) \\
12878.9   & **  & **  & **  & --         & H$_2$O \\
12879.8   & --  & --  & --  & ***(spl)   & Fe (VALD) \\
12880.4   & **  & *** & *   & --         & H$_2$O, FeH(w), CrH(w) \\
12883.5   & **  & *** & *** & **         & FeH, TiO(wk) \\
12884.4   & **  & *** & **(bl) & --      & FeH(w), H$_2$O(wk)\\
12885.3   & --  & **  & --  & --         & Ca (VALD) \\
12885.9   & *   & **  & *   & --         & FeH(w), H$_2$O \\
12888.8   & *   & *   & *   & *          & FeH, H$_2$O \\
12891.0   & *   & *   & *   & *          & FeH, H$_2$O \\
12892.4   & **  & *** & *** & **         & H$_2$O, FeH, TiO(wk) \\
12893.6   & *   & *   & *   & --         & CrH(w), FeH(w) \\
12897.0   & *   & **  & **  & *          & H$_2$O(?)\\
12897.7   & *   & *   & *   & *          & FeH, H$_2$O(?) \\
12899.8   & *** & *** & *** & ***        & Mn (VALD) \\
12902.6   & *   & **  & *   & *          & H$_2$O(?), TiO(wk) \\
12903.6   & *   & **  & *   & **         & FeH, H$_2$O(?) \\
12906.5   & **  & *** & **  & --         & H$_2$O, CrH(w), FeH(w), TiO(wk) \\
12907.3   & *   & **  & **  & *          & FeH, H$_2$O\\
12907.9   & --  & --  & --  & --         & Na (VALD) \\
12909.1   & *   & **  & -   & ***(spl)   & Ca, CrH, FeH, H$_2$O, TiO(wk)\\
12910.1   & --  & --  & --  & ***(spl)   & Cr (VALD) \\
12910.4   & **  & *** & *** & *(bl)      & FeH, H$_2$O\\
12913.5   & *   & *   & *   & --         & FeH, H$_2$O(?) \\
12916.5   & *** & *** & *** & **         & H$_2$O, FeH(w), TiO(w)\\
12917.3   & **  & ***(bl) & *** & *      & Na, H$_2$O, FeH(wk) \\
12920.0   & *   & *   & *   & ***(spl)   & Ti, H$_2$O, FeH(wk) \\
12922.1   & *   & **  & **  & **(spl)    & Cr, FeH \\
12923.9   & *   & *   & *   & *          & H$_2$O, FeH \\
12925.7   & **  & *** & **  & *          & FeH(?), H$_2$O(?)\\
12929.6   & **  & *** & **  & *          & FeH, H$_2$O \\
12932.5   & *** & *** & *** & **         & H$_2$O, FeH(w) \\
12934.6   & *   & *   & *   & *          & FeH, H$_2$O, CrH(w) \\
12937.0   & *   & *   & *   & ***(spl)   & Cr, H$_2$O \\
12939.5   & **  & *** & *** & *          & FeH \\
12941.7   & **  & **  & **  & **         & FeH(w), H$_2$O(?) \\
12943.8   & **  & *** & **  & *          & FeH, CrH(w), H$_2$O \\
12945.4   & *   & **  & **  & *          & FeH, H$_2$O(?) \\
12951.0   & **  & *** & *** & **         & CrH, H$_2$O \\
12952.7   & **  & **  & **  & **         & H$_2$O(?), FeH(w)\\
12953.8   & **  & *** & *** & *          & H$_2$O \\
12955.9   & **  & *** & *** & **         & H$_2$O, FeH(w) \\
12957.2   & **  & **  & *** & **         & H$_2$O \\
12959.6   & **  & *** & *** & **         & H$_2$O(?) \\
12962.3   & **  & **  & **  & **         & H$_2$O \\
12964.2   & *   & **  & **  & **         &  FeH(w), H$_2$O \\
12966.6   & **  & *** & *** & **         & H$_2$O, FeH(w) \\
12968.4   & *   & *   & **  & *          & H$_2$O \\
12969.9   & **  & *** & *** & *          & H$_2$O(?) \\
12973.9   & *   & --  & **  & *          & FeH \\
12975.2   & **  & **  & **  & *(bl)      & H$_2$O \\
12975.9   & **  & **  & **  & ***(spl)   & Mn, H$_2$O \\
12977.5   & *   & **  & **  & *          & H$_2$O, FeH(wk) \\
12978.7   & *   & **  & **  & **         & H$_2$O(?), FeH(w)\\
12980.8   & *** & *** & *** & *          & CrH, H$_2$O, FeH(w)\\
12982.2   & *   & *   & *   & *          & H$_2$O(?)\\
12988.9   & *   & **  & *   & *          & H$_2$O \\
12990.8   & *   & *** & **  & --         & FeH, H$_2$O\\
12993.2   & *   & *   & *   & *          & H$_2$O, CrH(w), FeH(w)\\
12996.3   & *   & **  & *   & *          & H$_2$O(?)\\
12998.4   & **  & *** & **  & *          & FeH, H$_2$O \\
13000.1   & *   & *** & *   & *          & H$_2$O(?), FeH(?) \\
13000.7   & *   & *** & *   & --         & H$_2$O \\
13001.4   & --  & --  & --  & **(spl)    & Ca \\
13001.9   & *   & *** & *   & **         & H$_2$O \\
\hline
\multicolumn{6}{l}{asterisks correspond to different spectral line intensities:}\\
\multicolumn{6}{l}{* - 0.96-0.99, ** - 0.91-0.95, *** - $<$0.95 (see text) }\\
\multicolumn{6}{l}{bl - the line feature is blended by other features}\\
\multicolumn{6}{l}{w - the wing of the line}\\
\multicolumn{6}{l}{wk - (very) weak line}\\
\multicolumn{6}{l}{? - this species probably forms the line-feature}\\
\multicolumn{6}{l}{spl - the line is split by magnetic field (solar lines)}\\
\multicolumn{6}{l}{VALD - Vienna Atomic Line Database (Kupka et al. 1999)}\\
\multicolumn{6}{l}{`--' - not identified}\\
\end{longtable}
}
\newpage

\newpage
\clearpage

\begin{table}[h]
\caption[]{Line identifications and strengths for the programme stars LP944-20 and 2M0345+25 are tabulated.}
\begin{tabular}{cccc}
\hline\hline
WL$\pm$0.3\AA, air &  LP944-20  & 2M0345+25 & Notes \\
\hline
12874.4 & ***     & ***     & FeH, H$_2$O(bl) \\
12877.7 & ***     & ***     & FeH(?), TiO(wk,?), H$_2$O(?)  \\
12880.4 & ***     & --(bl)  & H$_2$O, FeH(?)\\
12883.5 & ***     & --(bl)  & FeH, TiO(wk) \\
12884.4 & ***     & ***     & FeH(w), H$_2$O(wk)\\
12885.3 & ***     & ***     & Ca (VALD), FeH \\
12892.4 & ***     & **      & H$_2$O, FeH, TiO(wk) \\
12899.8 & ***     & ***     & Mn (VALD) \\
12906.5 & ***     & **      & H$_2$O, FeH(w), TiO(wk) \\
12907.9 & ***(bl) & **      & Na (VALD) \\
12910.4 & --      & --      &  CrH, H$_2$O, FeH\\
12916.5 & ***     & ***(bl) & H$_2$O, FeH(w), TiO(w)\\
12917.3 & ***     & ***     & Na, H$_2$O, FeH(wk) \\
12925.7 & ***     & *       & FeH(w), H$_2$O(?)\\
12929.6 & ***     & --      & FeH(w), H$_2$O(?) \\
12932.5 & ***(bl) & **(bl)  & H$_2$O, FeH(w) \\
12939.5 & ***     & *       &  FeH \\
12943.6 & --      & --      &  FeH, CrH, H$_2$O \\
12951.0 & ***     & -- (bl) &  CrH, H$_2$O, FeH \\
12953.8 & --      & **(bl)  & H$_2$O, FeH \\
12955.9 & ***     & --      & H$_2$O, FeH(w) \\
12957.2 & ***     & **(bl)  & H$_2$O \\
12959.6 & ***(bl) & **      & H$_2$O(?) \\
12966.6 & ***     & --      & H$_2$O, FeH(w) \\
12969.7 & ***     & ***     & H$_2$O \\
12980.8 & ***     & **      & H$_2$O, CrH, FeH \\
12990.8 & ***     & ***     & FeH, H$_2$O\\
12998.4 & ***(bl) & ***(bl) & FeH, H$_2$O, CrH(w) \\
13000.1 & ***     & ***(bl) & H$_2$O(?) \\
13000.7 & ***     & ***     & H$_2$O \\
13001.9 & ***     & ***     & H$_2$O \\
\hline
\multicolumn{4}{l}{asterisks correspond to different spectral line intensities:}\\
\multicolumn{4}{l}{* - 0.96-0.99, ** - 0.91-0.95, *** - $<$0.95 (see text) }\\
\multicolumn{4}{l}{bl - the line feature is blended by other features }\\
\multicolumn{4}{l}{w - the wing of the line}\\
\multicolumn{4}{l}{wk - (very) weak line}\\
\multicolumn{4}{l}{? - this species probably forms the line-feature}\\
\multicolumn{4}{l}{spl - the line is split by magnetic field (solar lines)}\\
\multicolumn{4}{l}{VALD - Vienna Atomic Line Database (Kupka et al. 1999)}\\
\multicolumn{4}{l}{`--' - not identified}\\
\hline
\end{tabular}
\end{table}

\begin{table}[h]
\caption[]{J-band Mn~I lines basic parameters.}
\begin{tabular}{cccc}
\hline\hline
Wavelength, \AA & $\chi$, eV&log~gf, VALD &log~gf$^{\star}$, Me99\\
\hline
12899.764 & 2.114 & -1.059 & -1.76 \\
12975.912 & 2.889 & -0.937 & -1.57 \\
\hline
\multicolumn{4}{l}{VALD - Kupka et al. (1999); Me99 - Mel$\grave e$ndez (1999).}\\
\multicolumn{4}{l}{$\star$- value for the main hyperfine component.}\\
\end{tabular}
\end{table}

\end{document}